\documentclass[aps,twocolumn,prb,showpacs,superscriptaddress,floatfix]{revtex4}
\usepackage{graphics,amssymb,amsmath,epsfig}

\begin{document}

\title{Relaxation Dynamics and Interrupted Coarsening\\
      in Irrationally Frustrated Superconducting Arrays}

\author{Gun Sang Jeon}
\affiliation{Department of Physics and Astronomy,
Seoul National University, Seoul 151-747, Korea}

\author{Sung Jong Lee}
\affiliation{Department of Physics, University of Suwon, Kyonggi-do
445-743, Korea}

\author{Bongsoo Kim}
\affiliation{Department of Physics, Changwon National University,
           Changwon 641-773, Korea}

\author{M.Y. Choi}
\affiliation{Department of Physics and Astronomy, Seoul National
University, Seoul 151-747, Korea} \affiliation{Asia-Pacific Center
for Theoretical Physics, Pohang 790-784, Korea}

\begin{abstract}
Equilibrium and non-equilibrium relaxation behaviors of
two-dimensional superconducting arrays are investigated via
numerical simulations at low temperatures in the presence of
incommensurate transverse magnetic fields, with frustration
parameter $f= (3-\sqrt{5})/2$. We find that the non-equilibrium
relaxation, beginning with random initial states quenched to low
temperatures, exhibits a three-stage relaxation of chirality
autocorrelations. At the early stage, the relaxation is found to
be described by the von Schweidler form.
Then it exhibits power-law behavior in the intermediate time scale
and faster decay in the long-time limit, which together can be
fitted to the Ogielski form;
for longer waiting times, this crosses over to a stretched
exponential form.
We argue that the power-law behavior in the intermediate time
scale may be understood as a consequence of the coarsening
behavior, leading to the local vortex order corresponding to
$f=2/5$ ground-state configurations. High mobility of the vortices
in the domain boundaries, generating slow wandering motion of the
domain walls, may provide mechanism of dynamic heterogeneity and
account for the long-time stretched exponential relaxation
behavior. It is expected that such meandering 
fluctuations of the low-temperature structure give rise to finite
resistivity at those low temperatures; this appears consistent
with the zero-temperature resistive transition in the limit of
irrational frustration.

\end{abstract}

\pacs{74.50+r, 67.40.Fd}


\maketitle

\section{Introduction}

Relaxation properties of systems with a great number of metastable
states have attracted much attention in the recent
decades.\cite{review,2DSG,3DSG,aging,Aging,CG} These systems are
usually characterized by the existence of {\em both} disorder and
frustration. A well-known system is the Ising spin glass
model,\cite{review} which displays non-exponential relaxation at
low temperatures.\cite{2DSG,3DSG} Interesting aging
phenomena\cite{aging} were also observed in this
system.\cite{Aging} The Coulomb glass model, which is another
interesting disordered system, has recently been shown to exhibit
relaxation of the stretched exponential form.\cite{CG} Such
non-exponential relaxation was also reported in frustrated systems
without disorder and its relation with the percolation transition
was discussed.\cite{Frustrated}

In a superconducting array, frustration can be induced in a controllable
way by applying an external magnetic field. It has crucial effects on
thermodynamics of the system and results in a variety of equilibrium
properties.\cite{TJ,arrayreview} Such remarkable diversity in the equilibrium
properties naturally leads to expectations that a variety of interesting dynamic
behaviors is also present, particularly in relaxation toward equilibrium.

An interesting limiting situation arises in the presence of
irrational frustration, the most typical case of which is provided
by the frustration parameter $f = 1- g$ with the golden number
$g\equiv (\sqrt{5}-1)/2$. The system was first suggested to
exhibit a spin-glass-like phase at low temperatures due to
self-generated disorder.\cite{ifxy_hal} Subsequently, it was
argued to display novel finite-size effects\cite{IFXY_finite_size}
that the size-dependent transition temperature decreases
monotonically with the system size, resulting in the absence of a
finite-temperature transition.\cite{IFXY1} Simulation results for
the current-voltage characteristics supported the zero-temperature
transition,\cite{IV_ifxy_Granato} while experimental results were
interpreted to exhibit finite-temperature
transitions.\cite{IV_wire_net,IV_ifxy_skku} Recent Monte Carlo
simulations, investigating the vortex configuration at low
temperatures in the system with $f$ given by rational approximants
to $1-g$, indicate the existence of a low-temperature phase where
the helicity modulus takes a finite value 
along one direction and vanishes along the other down to very low
temperatures.\cite{Gupta, ifxy_conf, kolahchi2}
%
From dynamical perspectives, due to the existence of many
metastable states that are almost degenerate with one another
(which is attributable to the incommensurate magnetic field), one
may expect characteristic slow 
relaxation in the array with irrational frustration. 
Dynamic simulations of this system, based on simple Langevin
dynamics, indeed disclosed a crossover temperature below which
strongly nonexponential relaxation emerges, exhibiting some
analogy to the behavior of supercooled liquids.\cite{bk_sjl_97}

In this paper we investigate the relaxation behavior of the
superconducting array with irrational frustration in both
equilibrium and nonequilibrium situations, employing the
resistively-shunted-junction (RSJ) dynamics in the overdamped
limit (i.e., junction capcitances are neglected). Note that the
present RSJ dynamics can, in principle, be realized in real
junction array experiments; this is in contrast to the Langevin
dynamics employed in existing simulations, which assumes
hypothetic dissipation between the superconducting islands and the
ground.

At equilibrium, we observe that chirality autocorrelations are
characterized by stretched exponential relaxation (with a
temperature-dependent stretching exponent) in a wide range of
intermediate and low temperatures. The relaxation time exhibits
non-Arrhenius behavior with the Vogel-Tammann-Fulcher type of
temperature dependence.

On the other hand, beginning with random initial states quenched
to low temperatures ($T < 0.14$), the dynamics exhibits slow aging
behavior, not reaching the equilibrium relaxation within our
computing time. It is observed that, for a short time
(approximately up to time $t \approx 10$ to $20$ depending on the
waiting time and temperature), relaxation of the chirality
autocorrelation function exhibits the von Schweidler
behavior.\cite{Schweid}
After the short time, chirality autocorrelations for short waiting
times relax according to the so-called Ogielski form
with temperature-dependent exponents. 
For longer waiting times, this behavior slowly changes into the stretched
exponential form 
with a temperature-dependent stretching exponent. 
The Ogielski form of the nonequilibrium relaxation is
characterized by power-law behavior in the intermediate time scale
and faster decay in the long-time region. The emergence of
power-law behavior in the intermediate time scale suggests the
presence of a sort of coarsening dynamics with dynamic scaling,
which is supported by the evolution of vortex patterns.

This is also consistent with the vortex configurations in
low-energy states obtained from a global optimization algorithm
such as conformational space annealing (CSA).\cite{CSA,csa_work}
Those vortex configurations obtained from CSA exhibit interesting
features: There exist approximately parallel domains of local
vortex order corresponding to the staircase ground state of
$f=2/5$.\cite{staircase} Those domains are separated by
domain-wall regions, consisting of characteristic local
arrangements of four consecutive vacancies along one (horizontal
or vertical) direction, thereby neighboring domains are
parallel-shifted by two horizontal (or vertical) lattice units.
Reminiscent of the smectic order in a liquid crystal, such
configurations may be described as smectic (liquid-crystalline)
arrangements of diagonal chains of vortices. Similar features of
the vortex lattice were also reported
for the case of frustration $f=13/34$ and $21/55$, which are
rational approximants to the irrational value $f =
1-g$.\cite{kolahchi2}

Evolution of the vortex pattern in general exhibits growth of
local vortex order with time toward the low-temperature
anisotropic state described above. This power-law behavior of
coarsening is expected to be interrupted by long-time fluctuations
of domain walls, limiting the maximum local domain size to around
$12$ lattice units on average.
In the long-time limit, vortex motions occur predominantly in the
domain-wall region through transfer of vortices between
neighboring diagonal vortex chains, which correspond also to the
motion of four consecutive vacancies. Accordingly, the higher
mobility of the vortices located in domain-wall regions can
naturally explain dynamic heterogeneity in the system, leading
also to the stretched exponential relaxation.

As for the resistive transition, one can argue that, as long as
the domain-wall defects keep fluctuating with unbounded
displacement, the system would remain resistive (i.e., exhibiting
finite resistance) even at low temperatures below the transition
to the anisotropic phase, since there always exists a direction
along which the helicity modulus vanishes. Those domain-wall
defects are likely to freeze at a much lower temperature which
vanishes in the limit of irrational frustration; thus concluded is
a zero-temperature resistive transition in the system with
irrational frustration. Note, however, that this argument applies
to a pure system with no quenched disorder. In a real
Josephson-junction array, disorder is inevitable in the
distribution of critical currents of individual junctions, which
leads to pinning of domain-wall defects at finite temperatures.
This may explain the recent experimental results reporting a
finite-temperature resistive transition near incommensurability of
the magnetic frustration.\cite{IV_ifxy_skku}


This paper is organized as follows: In Sec. II we introduce the equations
of motion for the RSJ dynamics of the system in the fluctuating twist
boundary conditions. Section III presents the results of simulations performed
on the equations of motion. Both equilibrium and nonequilibrium relaxation behaviors
of chirality autocorrelations are examined and coarsening, interrupted by fluctuating
domain-wall defects, is addressed. Finally, a summary is given in Sec. IV.

\section{Equations of Motion}

We begin with the set of equations of motion for the phases
$\{\phi_i\}$ of the superconducting order parameters in an
$L\times L$ square array. In the RSJ model under the fluctuating
twist boundary conditions,\cite{FTBC} they read:
\begin{equation} \label{eq:dyn1}
{\sum_j}' \left[
\frac{d \widetilde{\phi}_{ij}}{dt}
+ \sin (\widetilde{\phi}_{ij}-{\bf r}_{ij} \cdot {\bf \Delta})
+ \zeta_{ij}\right]=0 ,
\end{equation}
where we have employed the abbreviations
$\widetilde{\phi}_{ij}\equiv \phi_i{-}\phi_j{-}A_{ij}$ and ${\bf
r}_{ij}\equiv {\bf r}_i{-} {\bf r}_j$, and the primed summation
runs over the nearest neighbors of grain $i$. The position of
grain $i$ is represented by ${\bf r}_i = (x_i , y_i )$ with the
lattice constant set equal to unity while the gauge field $A_{ij}$
is given by the line integral of the vector potential ${\bf A}$:
\begin{equation}
A_{ij} \equiv \frac{2\pi}{\Phi_0} \int_{{\bf r}_i}^{{\bf r}_j}
{\bf A}\cdot d{\bf l}
\end{equation}
with the flux quantum $\Phi_0 \equiv hc/2e$. The frustration
parameter $f$, which measures the number of flux quanta per
plaquette, is given by the directional sum of the gauge field
$A_{ij}$ around a plaquette:
\begin{equation} \label{eq:sumA}
f \equiv \frac{1}{2\pi}\sum_P A_{ij} .
\end{equation}

In Eq.~(\ref{eq:dyn1}) the energy and the time have been expressed
in units of $\hbar I_c / 2e$ and $\hbar/2eRI_c$, respectively,
with single-junction critical current $I_c$ and shunt resistance
$R$. The thermal noise current $\zeta_{ij}$ is assumed to be white,
satisfying
\begin{equation}
\langle \zeta_{ij}(t{+}\tau) \zeta_{kl} (t) \rangle
= 2 k_B T  \delta (\tau) (\delta_{ik}\delta_{jl} {-}
\delta_{il}\delta_{jk})
\end{equation}
at temperature $T$.
Henceforth we set the Boltzmann constant $k_B \equiv 1$, thus measuring the
temperature in units of $\hbar I_c /2ek_B$.
The dynamics of the twist variables
${\bf \Delta} \equiv (\Delta_x, \Delta_y )$ is governed by
\begin{equation} \label{eq:dyn2}
\frac{d\Delta_{a}}{dt}
=\frac{1}{L^2}
\sum_{\langle ij \rangle_{a}}
\sin(\widetilde{\phi}_{ij}-\Delta_{a})
+ \zeta_{a} ,
\end{equation}
where $\sum_{\langle ij \rangle_{a}}$ denotes the
summation over all nearest-neighboring pairs in the $a$-direction $(a = x, y)$
and $\zeta_a$ satisfies
\begin{equation}
\langle \zeta_a(t{+}\tau) \zeta_b (t) \rangle
=  \frac{2T}{L^2} \delta_{ab} \delta (\tau).
\end{equation}

To study the relaxation of the system, we let the system evolve from
some random initial configurations quenched to given temperatures and
measure the chirality autocorrelation function:
\begin{equation}
C_q (t{+}t_w, t_w ) \equiv \frac{1}{L^2 f (1-f)} \sum_{\bf R}
\langle q_{\bf R}(t+t_w) q_{\bf R}(t_w) \rangle
\label{corr}
\end{equation}
with the waiting time $t_w$.
Here the chirality is defined to be
\begin{equation}
q_R(t) \equiv \frac{1}{2\pi} \sum_P
\left[\widetilde{\phi}_{ij}(t)-{\bf r}_{ij} \cdot {\bf \Delta}(t)
\right],
\end{equation}
where $\sum_P$ denotes the directional plaquette summation of links
around dual lattice site {\bf R} and the phase difference
$\widetilde{\phi}_{ij}(t)-{\bf r}_{ij} \cdot {\bf \Delta}(t)$
is defined modulo $2\pi$ in the range $(-\pi,\pi]$.


In numerical simulations, we have integrated directly the
equations of motion (\ref{eq:dyn1}) and (\ref{eq:dyn2}) via the
modified Euler method with time step $\Delta t = 0.05$. The time
step has been varied, only to give no appreciable difference. We
have considered mostly systems of linear size $L=55$ and $89$,
taking averages typically over $100$ to $600$ ensembles with
random initial states. The size has been chosen as members of the
Fibonacci sequence, thus to minimize the boundary effects due to
the irrational frustration $f=1-g$.  For comparison, we have also
considered close rational approximants $f=21/55$ and $34/89$, to
find no qualitative difference in relaxation dynamics.



\section{Simulation Results}

\begin{figure}[htb]
\centerline{\epsfig{file=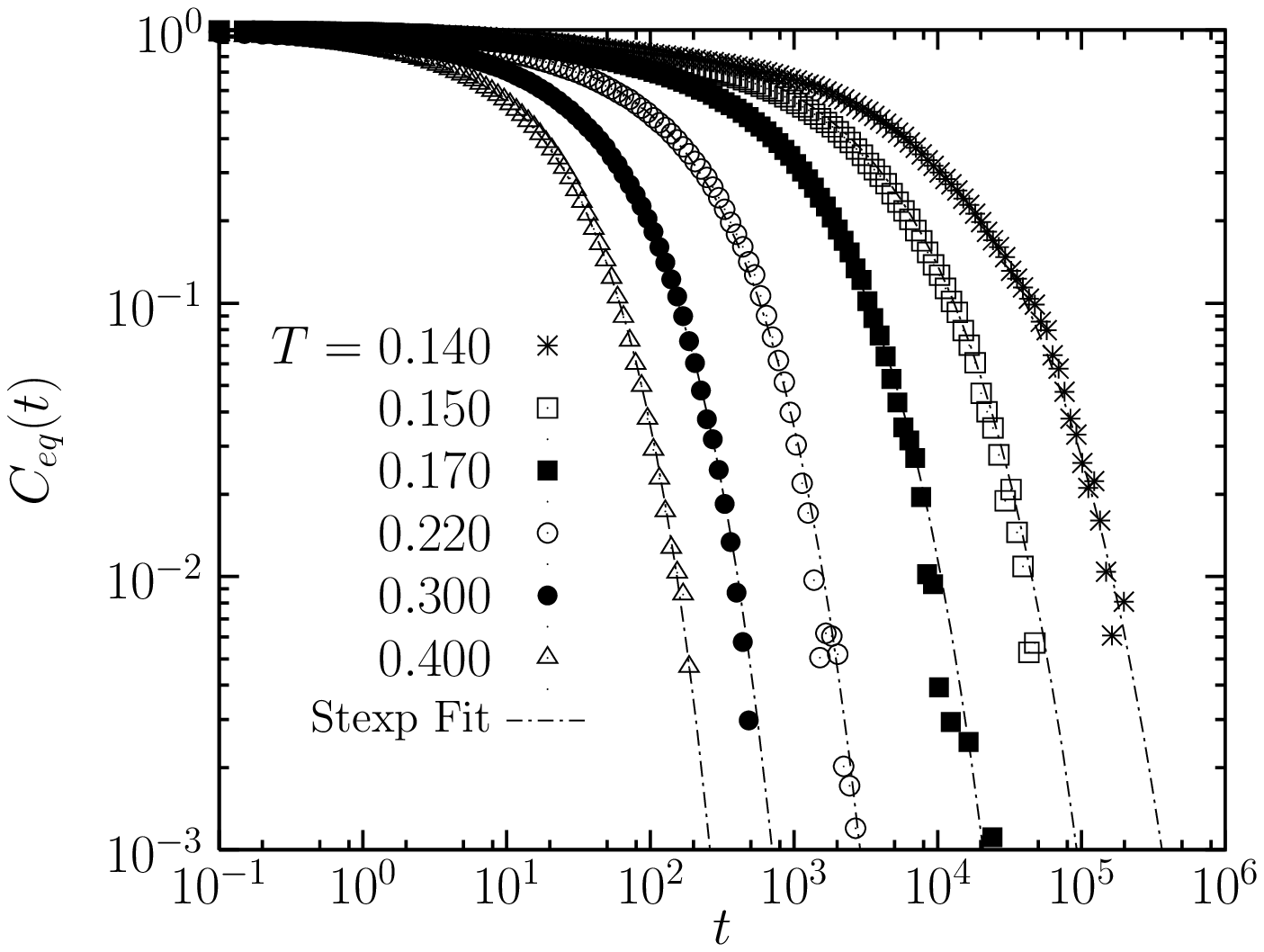,width=7cm}}
\centerline{(a)}
\vspace*{0.5cm}
\centerline{\epsfig{file=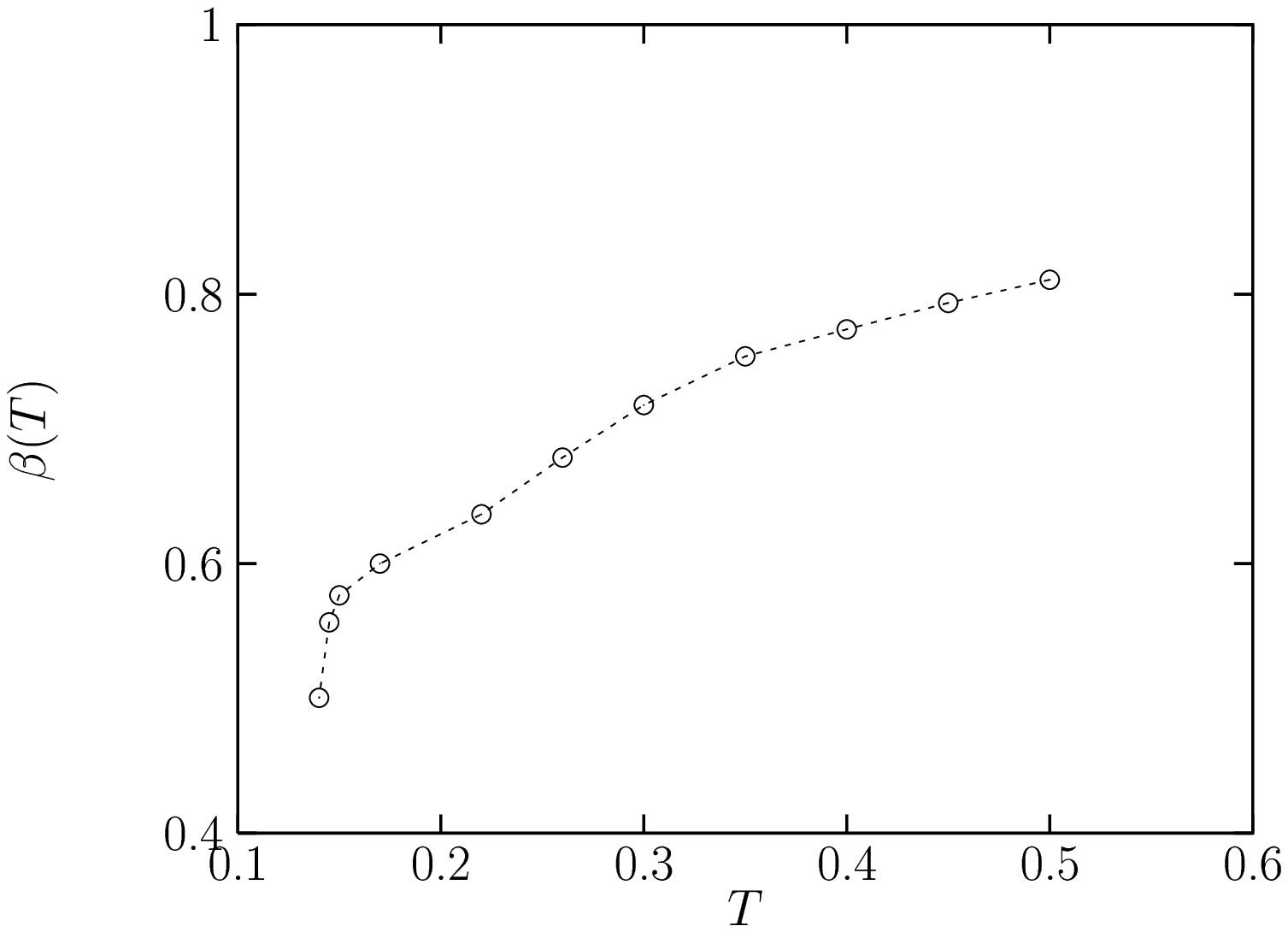,width=7cm}}
\centerline{(b)}
\vspace*{0.5cm}
\centerline{\epsfig{file=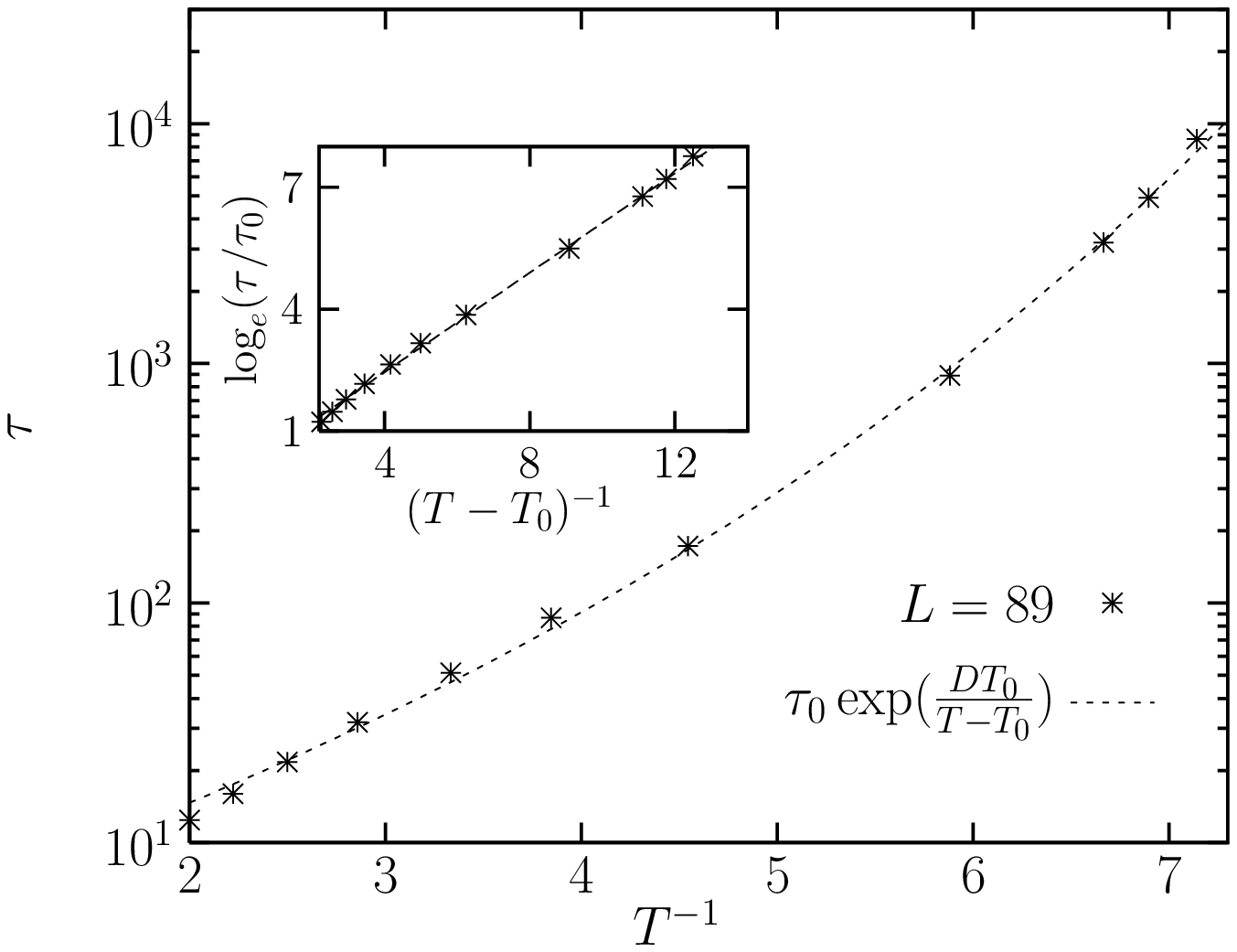,width=7cm}}
\centerline{(c)}
\caption{(a) Relaxation of the chirality autocorrelation function
(at equilibrium) $C_{eq} (t)$ at various temperatures $T$ in the
irrationally frustrated array of linear size $L=89$. Stretched
exponential fits are also plotted. (b) Stretching exponent versus
the temperature. The dotted line is merely a guide to the eye. (c)
Relaxation time versus the inverse temperature. The dashed line
represents the Vogel-Tammann-Fulcher fit with $T_0 = 0.06$ and
$\tau_0 =3.65$. The inset shows the result in a different scale.}
\label{fig:equil_relax_ifxy}
\end{figure}

We first consider the equilibrium relaxation behavior of the
system, obtained in the following way: Dynamic simulations are
performed with random initial states and then the autocorrelation
function in Eq.~(\ref{corr}) is computed for different values of
the waiting time. When the waiting time is sufficiently large (and
the temperature is not too low), the autocorrelation function no
longer depends on the waiting time, collapsing onto a single
relaxation function. This collapsed relaxation function is taken
to be the equilibrium relaxation function $C_{eq}(t)$. In this
way, equilibration of the system is achieved here, down to
temperature $T=0.14$.
In Fig.~\ref{fig:equil_relax_ifxy}(a), the behavior of $C_{eq}(t)$
is exhibited for several values of the temperature $T$. Excluding
the earliest time regime, one may fit this relaxation behavior to
a stretched exponential form: $C_{eq}(t) \approx  A \exp
[-(t/\tau)^{\beta}]$. Figure~\ref{fig:equil_relax_ifxy}(b) shows
that the stretching exponent $\beta$ decreases as the temperature
is lowered, reaching the value $\beta \approx 0.5$ at $T=0.14$. On
the other hand, Fig.~\ref{fig:equil_relax_ifxy}(c), plotting the
relaxation time $\tau$ versus the inverse temperature $T^{-1}$,
discloses the Vogel-Tamman-Fulcher behavior:
\begin{equation}
\tau (T) = \tau_0 \exp \left[D\frac{T_0}{T - T_0}\right]
\end{equation}
with the fragility parameter $D = 10.2$ and other parameters
$\tau_0 = 3.65$ and $T_0 = 0.06$. (Here $T_0$ is merely a fitting
parameter, perhaps not associated with a transition.) These
results are consistent with those from Langevin
dynamics.\cite{bk_sjl_97}

\begin{figure}[htb]
\centerline{\epsfig{file=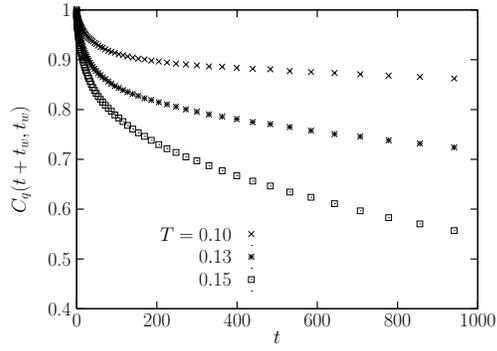,width=7cm}}
\centerline{(a)}
\vspace*{0.5cm}
\centerline{\epsfig{file=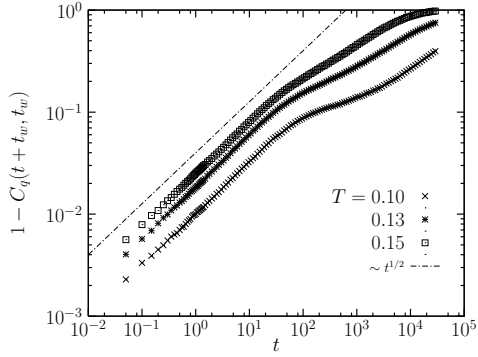,width=7cm}}
\centerline{(b)}
\vspace*{0.5cm}
\centerline{\epsfig{file=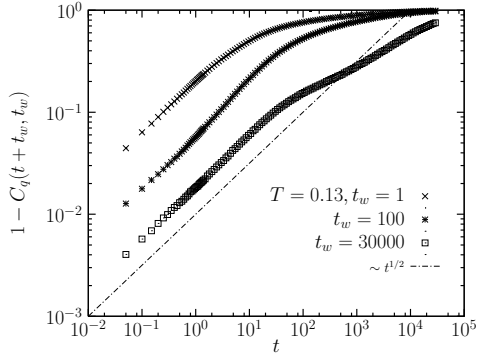,width=7cm}}
\centerline{(c)}
\caption{(a) Relaxation of the chirality autocorrelation function
$C_{q}(t{+}t_w , t_w )$ with time $t$ in the early-time regime at
temperatures $T = 0.10, 0.13$, and $0.15$ for $L=89$. The waiting
time is chosen to be $t_w =30000$. (b) The log-log plot of
$1-C_{q} (t{+}t_w , t_w )$ versus $t$ for the same data. For
almost three decades in the early-time regime, the relaxation
behavior is observed to follow $1-C_{q} (t{+}t_w , t_w ) \sim
t^{1/2}$ at all the three temperatures, as indicated by the
dot-dashed line. (c) The log-log plot of $1-C_{q} (t{+}t_w , t_w
)$ versus $t$ at temperature $T=0.13$ for three different waiting
times $t_w =1.0, 100$ and $30000$. It is shown that the early-time
behavior does not depend qualitatively on the waiting time.}
\label{fig:relax_short_tw0}
\end{figure}

As the temperature is further lowered, especially below $0.14$,
the system, starting from a random initial state, does not relax
to the equilibrium within the available computing time. Instead of
pursuing equilibrium relaxation at these low temperatures, we
probe nonequilibrium relaxation for various waiting times by
letting the system evolve from random initial states. It is found
that the resulting relaxation of chirality autocorrelations
proceeds in three stages. Figure~\ref{fig:relax_short_tw0}(a)
shows the time evolution of $C_q (t+t_w , t_w )$ for the waiting
time $t_w =30,000$ at temperature $T=0.15, 0.13$, and $0.10$. For
the same data, plotted in Fig.~\ref{fig:relax_short_tw0}(b) is
$1-C_q (t+t_w , t_w )$ versus time $t$, where one can see that,
for almost three decades beginning from the earliest time, the
relaxation fits nicely to the von Schweidler form $1- (t/\tau_0
)^b$ with $b \approx 1/2$. In Fig.~\ref{fig:relax_short_tw0}(c) we
show $1-C_q (t+t_w , t_w )$ versus time $t$ at temperature
$T=0.13$ for various waiting times. It is observed that the von
Schweidler behavior with $b \approx 1/2$ is rather robust, hardly
depending on the waiting time or the target temperature. This is
in contrast with the results of Langevin simulations, where the
value of $b$ tends to deviate from $1/2$, getting smaller at low
temperatures.\cite{bk_sjl_97} We presume that this discrepancy
arises from the difference in vortex dynamics (and diffusion) at
short times between RSJ dynamics and simple Langevin dynamics.

\begin{figure}
\centerline{\epsfig{file=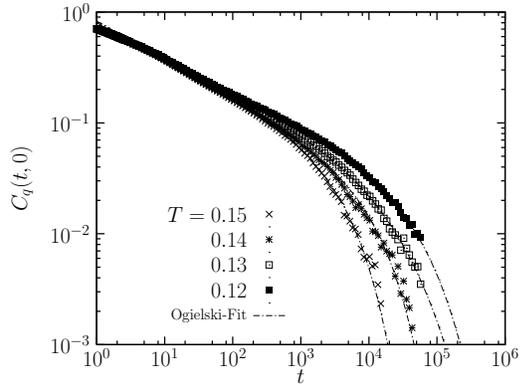,width=7cm}}
\caption{Nonequilibrium relaxation of the chirality
autocorrelation function $C_q (t{+}t_w , t_w)$ for zero waiting
time ($t_w =0.0$), starting from random initial states, at
temperature $T = 0.15, 0.14, 0.13$ and $0.12$. While in the
intermediate-time regime the nonequilibrium relaxation function
exhibits features of power-law behavior, this behavior is
interrupted by faster decay in the long-time regime, which
together can be fitted to the Ogielski form, as shown by the
dot-dashed lines.}
\label{fig:nonequil_zero_tw}
\end{figure}

\begin{figure}
\centerline{\epsfig{file=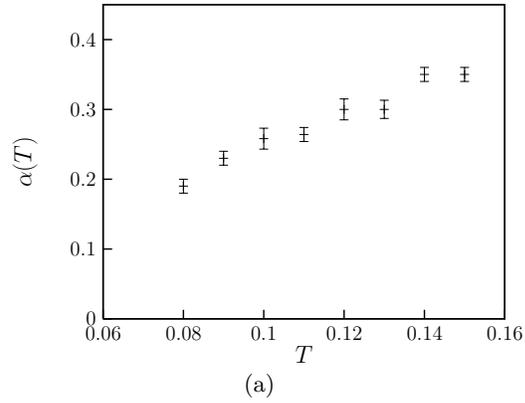,width=7cm}} \centerline{(a)}
\vspace*{0.5cm} \centerline{\epsfig{file=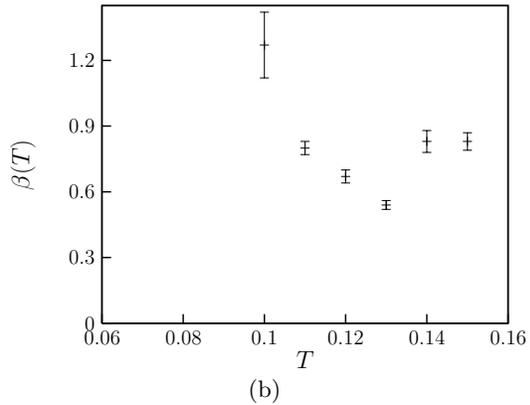,width=7cm}}
\centerline{(b)} \caption{Exponents (a) $\alpha$ and (b) $\beta$
versus temperature $T$ in the fit of the non-equilibrium
relaxation for $t_w = 0$ to the Ogielski form.}
\label{fig:nonequil_expo}
\end{figure}

We now turn to the relaxation behavior at intermediate and late
stages for various waiting times and temperatures.
Figure~\ref{fig:nonequil_zero_tw} shows the nonequilibrium
relaxation of the chirality autocorrelation function for null
waiting time at temperature $T= 0.15, 0.14, 0.13$, and $0.12$. It
is observed that the relaxation exhibits power-law behavior in the
intermediate-time regime, followed by faster decay in the
long-time regime. Here it is tempting to fit the relaxation of the
autocorrelation function to the Ogielski form:
\begin{equation} \label{eq:Ogielski}
C_q(t) \approx A_1 t^{-\alpha} \exp [-(t/\tau_1 )^\beta]
\end{equation}
with the exponents $\alpha$ and $\beta$ given in
Fig.~\ref{fig:nonequil_expo} and Table~I. Note that $\alpha $
depends substantially on the temperature, varying in the range of
$0.19$ to $0.35$ at temperatures between $0.08$ and $0.15$.

%




\begin{table}[ht]
\begin{tabular}{@{\hspace*{10mm}}  c @{\hspace*{15mm}}  c @{\hspace*{15mm}}  c
    @{\hspace*{10mm}}  } \hline \hline
\centering
  $T$ &  $\alpha$  & $\beta$   \\
\cline{1-3}
 0.08 &  0.19(1)   & -         \\
 0.09 & 0.23(1)    & -         \\
 0.10 & 0.258(15)  & 1.27(15)  \\
 0.11 &  0.264(10) & 0.80(3)   \\
 0.12 & 0.300 (15) & 0.67(3)   \\
 0.13 & 0.300 (13) & 0.54(2)   \\
 0.14 & 0.350(10)  & 0.83(4)   \\
 0.15 & 0.350(10)  & 0.83(4)   \\
\hline\hline
\end{tabular}
\caption{Exponents $\alpha$ and $\beta$ depending on the
temperature $T$. Note that $\beta$ is not shown for the cases of
$T=0.08$ and $0.09$. At these low temperatures, relaxation is too
slow to observe clearly the late-time stretched exponential part
within the computational time window, making it formidable to
estimate $\beta$.}
\label{t1}
\end{table}

\begin{figure}
\centerline{\epsfig{file=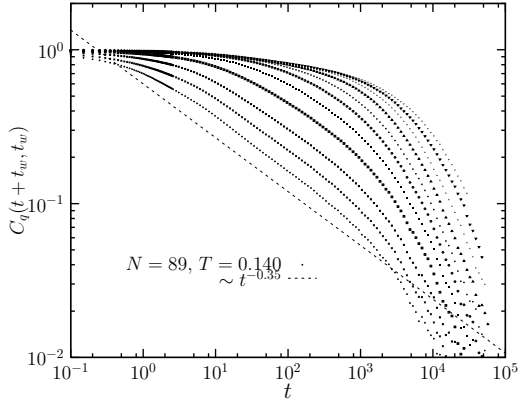,width=7cm}} \centerline{(a)}
\vspace*{0.5cm} \centerline{\epsfig{file=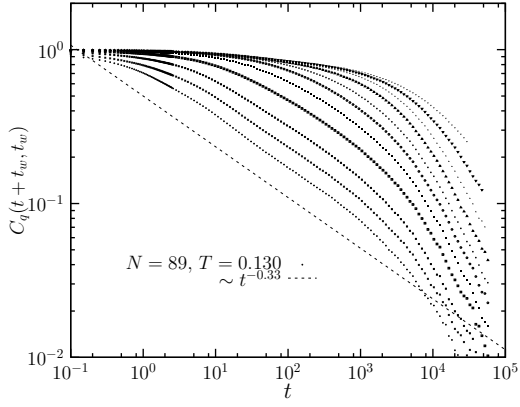,width=7cm}}
\centerline{(b)} \vspace*{0.5cm}
\centerline{\epsfig{file=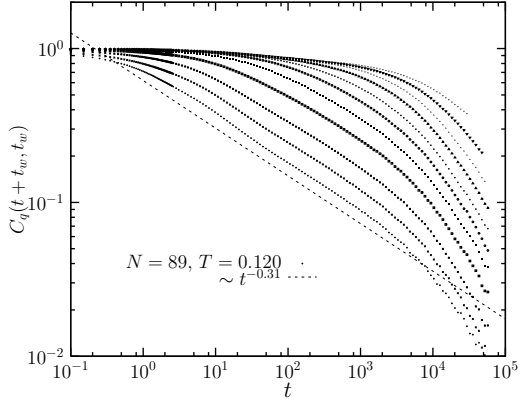,width=7cm}} \centerline{(c)}
\caption{Aging relaxation of the chirality autocorrelation
function $C_q (t{+}t_w , t_w)$ for various values of the waiting
time $t_w = 0, 1, 3, 10, 30, 100, 300, 1000, 3000, 10000$, and
$30000$ (from left to right), starting from random initial states,
at temperature $T = $ (a) 0.14, (b) 0.13, and (c) 0.12. Power-law
behavior $t^{-\alpha}$ in the intermediate-time regime is
manifested by dotted lines with the exponent $\alpha =$ (a) 0.35,
(b) 0.33, and (c) 0.31. As the waiting time grows, the relaxation
develops simple stretched exponential behavior.}
\label{fig:nonequil_aging_tw}
\end{figure}

\begin{figure}
\parbox{0.22\textwidth}{
\centerline{\epsfig{file=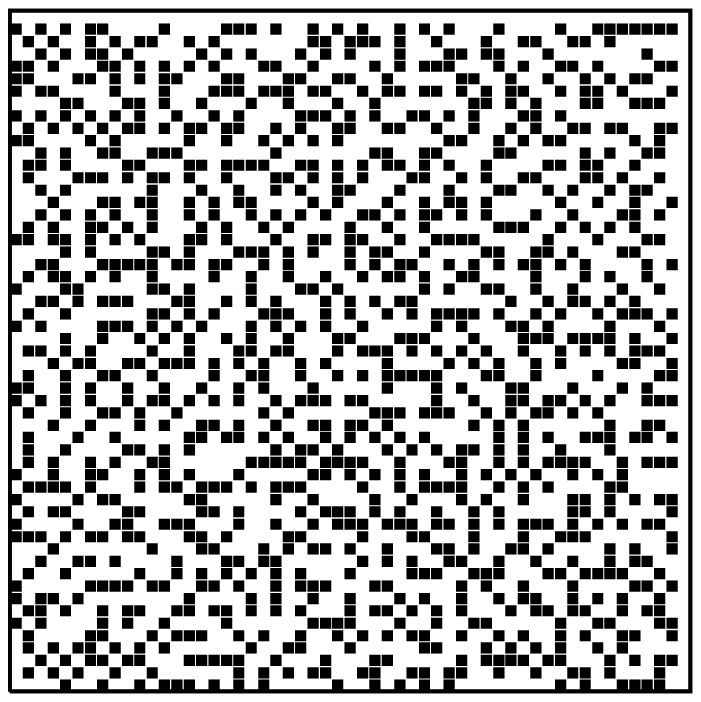,width=3.5cm}}
\centerline{(a)}
\vspace*{0.5cm}
\centerline{\epsfig{file=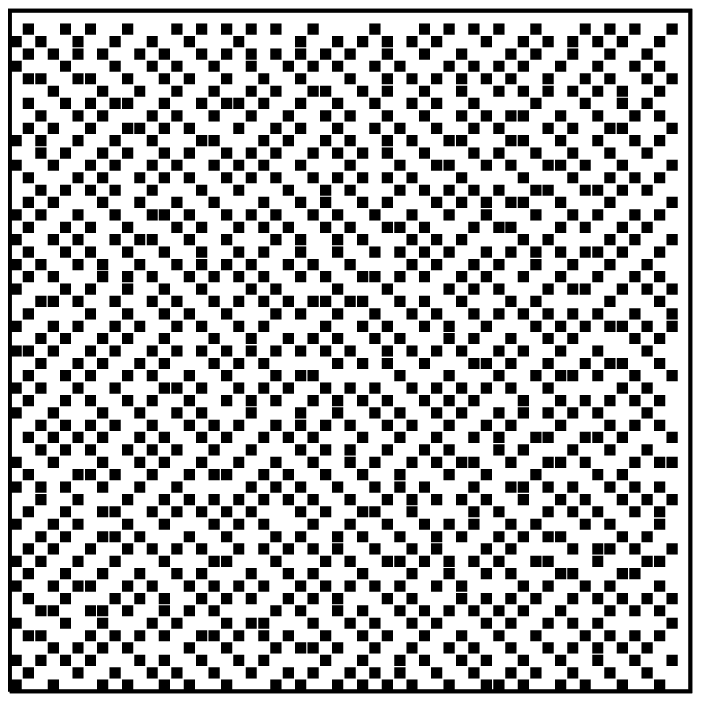,width=3.5cm}}
\centerline{(c)}
}
\hspace{0.02\textwidth}
\parbox{0.22\textwidth}{
\centerline{\epsfig{file=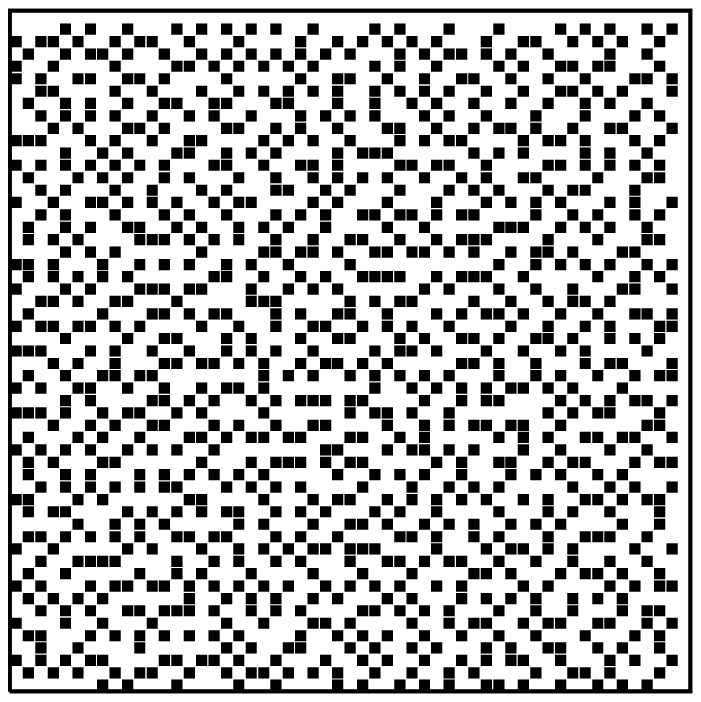,width=3.5cm}} \centerline{(b)}
\vspace*{0.5cm} \centerline{\epsfig{file=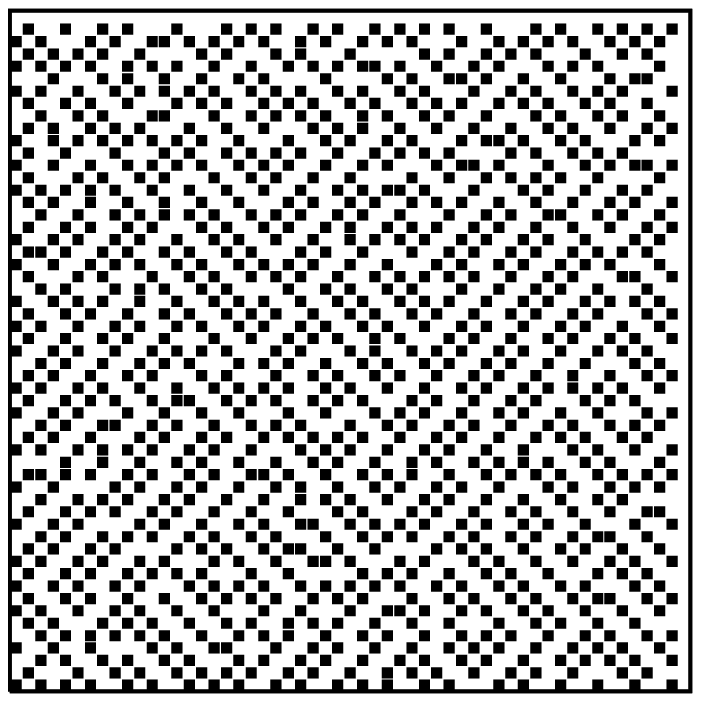,width=3.5cm}}
\centerline{(d)} } \caption{Snapshots of the vortex configuration
in the system of size $L= 55$, beginning from a random initial
state and quenched to temperature $T=0.130$, at time $t = $ (a)
0.0, (b) 8.0, (c) 128, and (d) 2048.} \label{fig:conf_growth_T013}
\end{figure}

Also shown in Fig~\ref{fig:nonequil_aging_tw} is the
nonequilibrium relaxation of the chirality autocorrelation
function for various waiting times, at temperature $T= 0.14,
0.13$, and $0.12$. For longer waiting times ($t_w > 100.0$), this
behavior crosses over to the stretched exponential form $C_q(t)
\approx A_2 \exp [-(t/\tau_2 )^\beta]$. Emergence of the power-law
behavior of the nonequilibrium relaxation at intermediate times
strongly suggests that there exist some coarsening processes in
the system. Figure~\ref{fig:conf_growth_T013} exhibits snapshots
of the vortex configuration at temperature $T =0.13$, taken at
several time instants; one can recognize slow growth of local
order corresponding to $f=2/5$ vortex patterns, with diagonal
chain structures. Since the frustration of the system is given by
$f=1-g$, which is slightly smaller than $2/5$, these locally
ordered domains of $f=2/5$ patterns may not grow to span the whole
system. Instead, there should exist finite length (and also time)
scales for the growth of these local domains, beyond which the
growth is interrupted by domain-wall regions of lower vortex
density, so that the net vortex density of the whole system
becomes equal to $1-g$.

\begin{figure}[htb]
\centerline{\epsfig{file=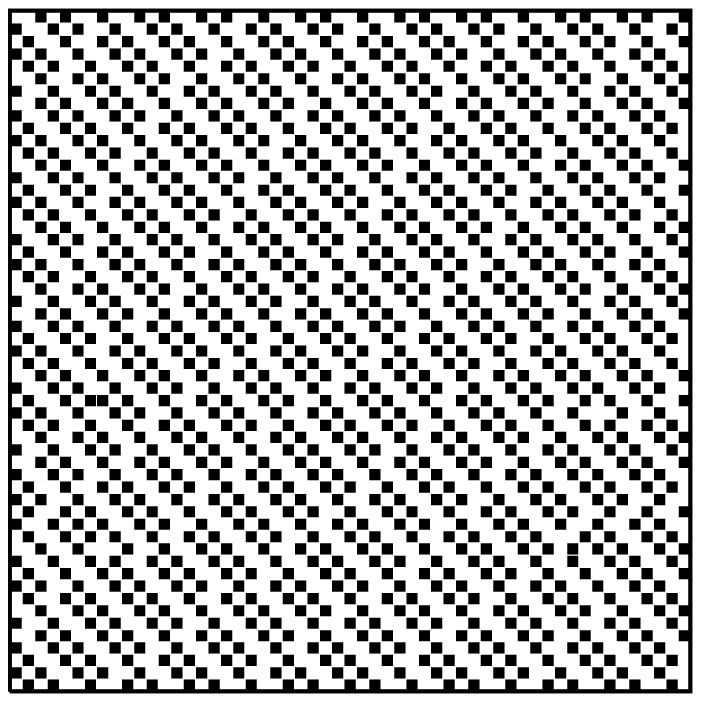,width=7cm}}
\centerline{(a)}
\vspace*{0.5cm}
\centerline{\epsfig{file=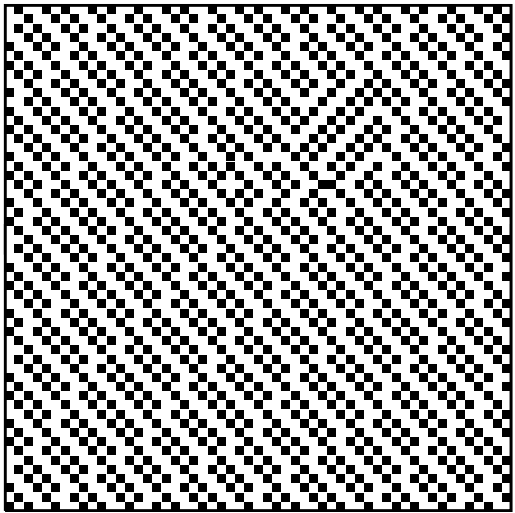,width=7cm}}
\centerline{(b)}
\caption{Snapshots of the vortex configuration obtained via CSA
for lattice size $L= 55$ with energy $E=$ (a) -1.273786 and (b)
-1.274077 (in units of $\hbar I_c /2e$), respectively.}
\label{fig:conf_csa_L55}
\end{figure}

In order to understand the vortex configuration attained in the
long-time limit at low temperatures, we investigate the
configuration of low-energy states by means of the efficient
optimization algorithm, CSA.\cite{CSA} Snapshots of typical vortex
configurations obtained via CSA are shown in
Fig~\ref{fig:conf_csa_L55}, where we observe domains of locally
ordered vortex patterns corresponding to the staircase ground
state of the system with frustration (or vortex density) $f=2/5$.
These domains of typical width $9$ to $12$ lattice spacings are
separated by domain-wall (line defect) regions that consist of
characteristic local arrangements of four neighboring vacancies
(see bar-shaped regions, each with four consecutive empty
plaquettes); there the vortex density is lower than the locally
ordered regions (of local vortex density $2/5$) in such a way that
the net vortex density of the whole system is precisely equal to
$1-g$ (which is less than $2/5$). This configurations may also be
described as a liquid-crystal-type arrangement of diagonal chains
of vortices (of length $9$ to $12$), with the neighboring chains
of vortices shifted in the diagonal direction by about half the
length of the chains. Similar configurations were observed in the
case of rational approximants to $1-g$.\cite{kolahchi2}
%

In the long-time limit, we expect that the coarsening dynamics
will lead eventually to the locally ordered configuration with
domain walls, shown in Fig.~\ref{fig:conf_csa_L55}. Within a
locally ordered domain, vortices are almost rigid and resist
moving. In contrast, those vortices at ends of the chains are
easily put in motion, jumping into the bar-shaped vacancies in
domain-wall regions and thus joining another vortex chain. This in
turn gives rise to domain-wall fluctuations (or equivalently,
fluctuations in the length of the vortex chains), which would lead
slowly but ultimately to the complete restructuring of the local
vortex configuration. It is thus expected that dynamic
heterogeneity naturally emerges from the existence of distinctly
mobile vortices in the domain-wall regions.\cite{hetero}

We believe that such domain-wall fluctuations and restructuring of
the local vortex configuration can explain the absence of freezing
of the relaxation at low temperatures. Namely, the low-temperature
state is presumably of a liquid crystalline type: Even though
there exists orientational order in the chain-like arrangement of
vortices, those diagonal chains can {\em flow} (like a liquid) due
to the transfer of vortices between neighboring chains,
corresponding to the fluctuating motion of the bar-shaped
vacancies and resulting in finite resistivity. At a much lower
temperature one may expect complete freezing of the domain walls
to occur. The freezing temperature should depend on the
commensurability of the distribution of the vacancy defects and
the underlying background vortex lattice; this is directly related
to the rationality of $f=p/q$ (with $p$ and $q$ relatively prime
integers) and the freezing temperature is expected to vanish in
the irrational limit ($q \rightarrow \infty$).

It is of some interest to note the similarity to the behavior
found in the lattice coulomb gas with charge density near the
golden number.\cite{LLK_2002} Even though the detailed ordering
pattern is different, it was found numerically that there exist
two-step transitions in the lattice coulomb gas with $1/3<f<2/5$,
where the intermediate phase corresponds to anisotropic striped
charge ordering together with mobile charges within
partially-filled channels. There the lower transition corresponds
to the complete freezing of charges within partially filled
diagonal channels, which occurs at temperatures sensitively
dependent on the rationality of $f$ due to the commensurability
effects. In this study our system is the Josephson-junction array
in the limit of irrational frustration, and we thus expect that
the true vortex freezing would occur at zero temperature.

As long as the domain-wall defects fluctuate with unbounded
displacement, the system should remain resistive, exhibiting
finite resistance. These domain-wall defects are expected to
freeze at much lower temperature of the order $1/q$ for $f=p/q$;
this leads to the zero-temperature resistive transition in the
limit of irrational frustration. Note that this argument applies
for pure systems with no quenched disorder. In the case of real
Josephson-junction arrays, some type of disorder, e.g., in the
distribution of the critical currents of individual junctions is
unavoidable. Therefore pinning of the domain-wall defects can
easily occur at finite temperatures, having resistance vanish.
This may explain the recent experiment reporting a
finite-temperature resistive transition near incommensurability of
the magnetic frustration, where disorder in the critical currents
of individual junctions was noted to be up to
$15\,\%$.\cite{IV_ifxy_skku}


\section{Summary}

We have studied the relaxation behavior of the chirality
autocorrelation function in two-dimensional superconducting arrays
under irrational frustration at intermediate and low temperatures.
Both equilibrium relaxation and nonequilibrium relaxation have
been investigated via numerical simulations of RSJ dynamics.
Equilibrium relaxation dynamics reveals characteristic features of
the stretched exponential form with the Vogel-Tamman-Fulcher
dependence of the relaxation time.

Nonequilibrium relaxation at low temperatures, beginning with
random initial states exhibits interesting waiting-time
dependence: For short waiting times, the relaxation follows a
power-law behavior in the intermediate-time regime and faster
decay in the long-time regime, characterized together by the
Ogielski form with temperature dependent exponents. For longer
waiting times, this gradually crosses over to a stretched
exponential form. Further, in all cases of short and long waiting
times, the relaxation at early time stage fits nicely to the von
Schweidler form 
with exponent about $1/2$. 

It has been argued that the power-law behavior originates from
coarsening dynamics up to a certain length scale, with the local
vortex order corresponding to the ground state of $f=2/5$, found
to be consistent with the vortex configurations of low-energy
states. This coarsening, however, is interrupted by the presence
of domain-wall defects. Such chain-like domain-wall regions are
expected to provide possible mechanism of dynamic heterogeneity
and stretched exponential relaxation. Since the RSJ dynamics
adopted in this study can be realized experimentally, it would be
of interest to carry out experiment and compare the results.

\acknowledgments

This work was supported in part by the BK21 Project.
M.Y.C. also acknowledges a visitor grant from the CNRS, France and
thanks the Laboratoire de Physique Th\'eorique, Strasbourg, for
its kind hospitality during his stay.



\end{document}